\documentclass[mathleft]{an}
\usepackage{graphicx}
\usepackage{times}
\newcommand{\msun}{~{\rm M}_\odot}
\newcommand{\binf}{\mathcal{B}}
\newcommand{\binfm}{\mathcal{B}_{M_1}(M_1)}
\newcommand{\binfall}{\mathcal{B}_{\rm all}}

\newcommand{\fq}{f_q(q)}

\newcommand{\fmall}{f_{M,{\rm all}}(M)}

\newcommand{\mmin}{M_{\rm min}}

\begin{document}

\Pagespan{1}{4}
\Yearpublication{2008}%
\Yearsubmission{2008}%
\Month{---}%
\Volume{329}%
\Issue{9/10}%
\DOI{10.1002/asna.200811061}%

\title{Pairing mechanisms for binary stars}

\author{M.B.N. Kouwenhoven\inst{1}\fnmsep\thanks{Corresponding author:
  \email{t.kouwenhoven@sheffield.ac.uk}\newline}
\and A.G.A. Brown\inst{2}
\and S.P. Goodwin\inst{1}
\and S.F. Portegies Zwart\inst{3,4}
\and L. Kaper\inst{3}
}
\titlerunning{Pairing mechanisms for binary stars}
\authorrunning{M.B.N. Kouwenhoven et al.}
\institute{
Dept. of Physics and Astronomy, University of Sheffield, Hicks Building, Hounsfield Road, Sheffield S3 7RH, UK
\and 
Leiden Observatory, University of Leiden, P.O. Box 9513, 2300 RA Leiden, The Netherlands
\and
Astronomical Institute, University of Amsterdam, Kruislaan 403, 1098 SJ Amsterdam, The Netherlands
\and
Section Computer Science, University of Amsterdam, Kruislaan 403, 1098 SJ Amsterdam, The Netherlands
}

\received{---}
\accepted{---}
\publonline{---}


\abstract{%
  Knowledge of the binary population in stellar groupings provides important information about the outcome of the star forming process in different environments. Binarity is also a key ingredient in stellar population studies and is a prerequisite to calibrate the binary evolution channels. In these proceedings we present an overview of several commonly used methods to pair individual stars into binary systems, which we refer to as the {\em pairing function}. Many pairing functions are frequently used by observers and computational astronomers, either for the mathematical convenience, or because they roughly describe the expected outcome of the star forming process. We discuss the consequences of each pairing function for the interpretation of observations and numerical simulations. The binary fraction and mass ratio distribution generally depend strongly on the selection of the range in primary spectral type in a sample. These quantities, when derived from a binary survey with a mass-limited sample of target stars, are thus not representative for the population as a whole. 
}

\maketitle


\section{Introduction}

Most stars form as part of a binary or multiple system (e.g., Duquennoy \& Mayor 1991;
Mason et al. 1998; Goodwin \& Kroupa 2005; Kouwenhoven et al. 2005, 2007b; Goodwin et al. 
2007), and about 15\% of these are part of a multiple ($N \geq 3$) 
system (Tokovinin \& Smekhov 2002;
Correia et al. 2006; Tokovinin et al. 2006; Hu et al. 2008). In order to understand
the star formation process, it is of crucial importance to characterize the properties
of the binary population. In these proceeding we discuss several methods of pairing
stars into binary systems, and refer to these as {\em pairing functions}. 
The pairing functions described in these proceedings are used in literature for (i) a description of the observed binary population, (ii) initial conditions for $N$-body simulations, and (iii) a description of the outcome of hydrodynamical simulations or theoretical predictions of the star formation process. The pairing function determines the distribution of stellar masses. Variations of the {\em primordial} pairing function with environment thus imply variations of the initial mass function (IMF) and the star formation process.


\section{Method and terminology} \label{section:method}

Consider a stellar population consisting of single stars and binary systems. The pairing algorithm used to generate the binaries in such a stellar population is called the {\em pairing function} (Kouwenhoven et al. 2008). This pairing function could, for example, be random pairing of the binary components from the initial mass function (IMF). 

For a binary system consisting of two stars with masses $M_1$ and $M_2\leq M_1$, we refer to $M_1$ as the primary star and $M_2$ as the companion star. The mass ratio of the system is $q=M_2/M_1$ with $0 < q \leq 1$, and the total mass of the system is $M_T=M_1+M_2$. For simplicity, we neglect the presence of triple and higher-order systems.

Depending on the pairing function, the mass of the primary star or the total mass of each binary system is drawn from a {\em generating mass distribution} $f_M(M)$, such as the Kroupa (2001) mass distribution. A certain fraction $\binf$ of the stars is assigned a companion star. Depending on the pairing function, additional manipulations may performed, such as swapping of primary and companion stars such that $M_1 \geq M_2$, or the rejection of low-mass companions (see \S\,\ref{section:pairingfunctions}). Note that, due to these manipulations, the stellar mass distribution (of all stars: primary, companion, and single stars) changes. The resulting mass distribution (``IMF'') is therefore generally not equal to $f_M(M)$. The only exception is random pairing (RP), where both components of each binary system are independently drawn from $f_M(M)$.


\section{Several pairing functions} \label{section:pairingfunctions}

There are numerous ways to pair individual stars into binaries. Several straightforward pairing functions are discussed below (we refer to Kouwenhoven et al. (2008) for details).

\noindent
-- {\em Random pairing} ({\bf RP}). Both components of each binary are drawn from a generating mass distribution $f_M(M)$ and paired into a binary system, and swapped, if necessary, so that the primary is the most massive of the two (e.g., Piskunov \& Mal'Kov 1991; Mal'Kov \& Zinnecker 2001). RP is the simplest possible pairing function, as it is fully characterised by $f_M(M)$ and $\binf$. RP is the only pairing function for which the resulting stellar mass distribution $\fmall$ of all stars is equal to $f_M(M)$. Random pairing in a stellar population implies a chaotic formation process, where stars are paired into binaries irrespective of their mass.

\noindent
-- {\em Primary constrained random pairing} ({\bf PCRP}). The primary mass $M_1$ is drawn from $f_M(M)$. The companion star is also drawn from $f_M(M)$, with the additional constraint that it is less massive than the primary: $M_2 \leq M_1$. Note the subtle difference between RP and PCRP.

\noindent
-- {\em Primary-constrained pairing} ({\bf PCP}).  The primary mass $M_1$ is drawn from the mass distribution $f_M(M)$, and the mass ratio $q$ is drawn from a mass ratio distribution $f_q(q)$. Finally, the companion mass $M_2=qM_1$ is calculated. A possible origin of this pairing function comes from disk fragmentation. The mass of a circumstellar disk is related to the mass of the primary star.

\noindent 
-- {\em Split-core pairing} ({\bf SCP}). The total mass $M_T=M_1+M_2$ of the binary system is drawn from a generating mass distribution, and the mass ratio of the binary system is drawn from a mass ratio distribution. Subsequently, $M_T$ is split up into two stars with a mass $M_1=M_T(1+q)^{-1}$ for the primary and a mass $M_2=M_T(1+q^{-1})^{-1}$ for the companion star. The resulting binary system than has a total mass $M_T=M_1+M_2$ and a mass ratio $q=M_2/M_1$. A pairing function similar to SCP is expected from the theory of fragmenting gas clumps.

For pairing functions PCP and SCP an additional complication may occur. If a low mass ratio $q$ is drawn from $f_q(q)$, it is possible that the companion mass is smaller than a certain minimum mass $\mmin$. Such a minimum mass may originate from star formation theories. A minimum companion mass also originates from our definition of a binary system: a star with a companion less massive than the deuterium-burning limit (i.e., a planet) is considered a single star, implying $\mmin \approx 0.02 \msun$. Adopting this lower limit would also imply that stellar and brown dwarf companions have the same formation mechanism (see, e.g., Kouwenhoven et al. 2007a for a discussion). There are three methods of treating these low-mass companions:

\noindent
-- {\em Accept all companions} ({\bf PCP-I/SCP-I}). All companions are accepted, irrespective of their mass. In this case, even stars with a planetary companion are considered to be in a binary system.

\noindent
-- {\em Reject low-mass companions} ({\bf PCP-II/SCP-II}). All companion stars with a mass smaller than a certain minimum mass $M_{\rm min}$ are rejected, and the primary star becomes a single star. For this pairing function, stars with a low-mass (planetary) companion are not considered as binary systems. Note that the latter manipulation decreases the binary fraction, and increases the average mass ratio, particularly among low-mass binaries.

\noindent
-- {\em Redraw low-mass companions} ({\bf PCP-III/SCP-III}). When a companion with $M_2<M_{\rm min}$ is drawn, it is rejected, and a new companion star is drawn. This procedure is repeated until $M_2>M_{\rm min}$. This pairing mechanism ensures that stars with low-mass (planetary) companions are not present in the resulting population, and that the resulting binary fraction is equal to $\binf$. This manipulation increases the average mass ratio, particularly among low-mass binaries.

\begin{figure*}[t]
\includegraphics[width=\textwidth,height=!]{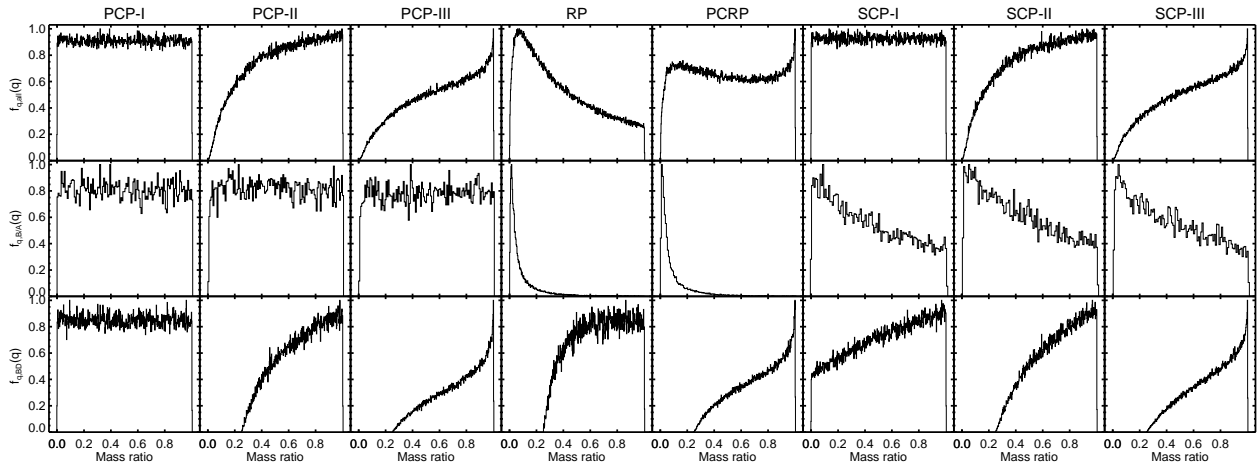}
\caption{
The mass ratio distributions resulting from the different pairing functions, for models with a Kroupa (2001) generating mass distribution, where applicable a generating mass ratio distribution $f_q(q)=1$, and a generating binary fraction of 100\%. From top to bottom, the panels show the overall mass ratio distribution (i.e., for all binaries), the mass ratio distribution for binaries with $1.5\msun \leq M_1 \leq 20\msun$, and for binaries with $0.02\msun \leq M_1 \leq 0.08\msun$. This figure illustrates that each pairing function results in a different overall and specific mass ratio distribution. Note that, apart from the differences in the mass ratio distributions, there are also differences in the resulting mass distribution and overall and specific binary fractions (see Fig.~\ref{figure:binaryfraction_versus_spectraltype}).
}\label{fig:massratiodistributions}
\end{figure*}


\section{Differences between pairing functions} \label{section:differences}

In this section we describe the main differences between the pairing functions described in \S\,\ref{section:pairingfunctions}. Most of the properties of these pairing functions  can be derived analytically. We refer to Kouwenhoven et al. (2008) for full derivations. 

For pairing function RP, the stellar mass distribution (of primaries, companions, and single stars) $\fmall$, is equal to the generating mass distribution $f_M(M)$. For all other pairing functions, $\fmall$ contains significantly more low-mass stars than $f_M(M)$. Only for SCP-I, the system mass distribution $f_{M_1,M_2,S}(M)$ is equal to $f_M(M)$. For pairing functions PCRP, PCP-I, PCP-II and PCP-III, the distribution of primary/single stars masses (i.e., the target stars in a survey for binarity) is equal to $f_M(M)$. 

The mass ratio distribution resulting from each pairing function is different. Moreover, for all pairing functions (except PCP-I) the mass ratio distribution varies with primary mass. This means that, for a given pairing function, the measured mass ratio distribution and binary fraction depends on the range in spectral type of the selected targets in a binary survey. An example of the mass ratio distribution for the nine different pairing functions is shown in Fig.~\ref{fig:massratiodistributions}, for three different targeted samples (all, high-mass, and low-mass target stars). For the other pairing functions, the overall mass ratio distribution is biased towards low values.

\begin{figure*}[t]
  \centering
  \includegraphics[width=0.9\textwidth,height=!]{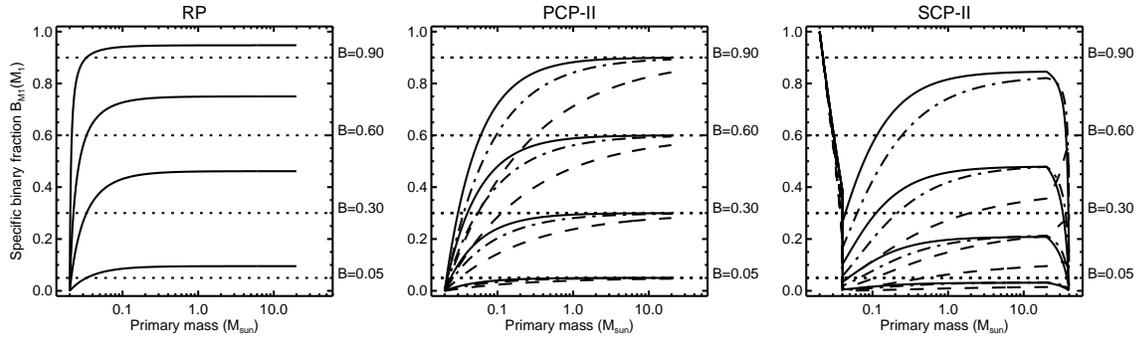} 
  \caption{The specific binary fraction $\binfm$ as a function of primary mass for pairing functions RP ({\em left}), PCP-II ({\em middle}) and SCP-II ({\em right}). Models with a generating binary fraction $\binf$ of (from bottom to top) 5\%, 30\%, 60\%, and 90\% are shown. The minimum companion mass is set to $0.02\msun$. The horizontal lines represent the generating binary fraction. No selection effects have been applied. {\em Left: } results for RP with the Salpeter mass distribution. {\em Middle} and {\em right:} results for PCP-II and SCP-II with mass ratio distributions $\fq\propto q^\gamma$, with $\gamma=0$ (solid curves), $\gamma=-0.3$ (dash-dotted curves), and $\gamma=-0.6$ (dashed curves). The figure shows that, even though we have applied the pairing function regardless of the primary spectral type, the binary fraction may vary strongly with primary spectral type.
    \label{figure:binaryfraction_versus_spectraltype} }
\end{figure*}

Each pairing function is characterised by a generating binary fraction $\binf$, the fraction of stars that is assigned a companion, prior to further manipulations associated with each pairing function. For most pairing functions the resulting overall binary fraction $\binfall$ is equal to $\binf$, except for PCP-II and SCP-II, where $\binfall < \binf$ due to the rejection of low-mass (planetary) companions. The binary fraction is a function of primary mass for RP (except when $\binf =100\%$), PCP-II and SCP-I/II/III. Examples on the varying binary fraction for these pairing functions are shown in Fig.~\ref{figure:binaryfraction_versus_spectraltype}.


\section{Implications and interpretation} \label{section:implications}

The outcome of star formation theories can be described using pairing functions. Disk fragmentation, for example, may result in a pairing function similar to PCP-II, while the fragmentation of pre-stellar cores may result in a population similar to that of SCP-II. After (and during) the star formation process, the pairing function may be altered due to gravitational interactions, as well as stellar evolution, which can alter the properties of binary systems.

The easiest method to distinguish observationally between the different pairing functions is to measure the mass ratio distribution as a function of primary spectral type. Using this method, pairing functions RP and PCRP have been ruled out by observations (see, e.g., Kouwenhoven et al 2008, and references therein). A realistic pairing function should be able to account for the large number of massive, equal-mass binaries (``twin binaries'') among massive stars (e.g., Pinsonneault \& Stanek 2006; Lucy 2006; S\"{o}derhjelm 2007), as well as the small number of brown dwarf companions among stellar-type objects (e.g., Kouwenhoven et al. 2007a).

Surveys for binarity have indicated that the binary fraction depends strongly on spectral type (see, e.g., Sterzik \& Durisen 2004; K\"{o}hler et al. 2006; Lada 2006; Bouy et al. 2006). For early-type stars (O/B/A) the binary fraction is close to 100\%. The binary fraction decreases to $50-60\%$ for F- and G-type stars, and decreases further to $30-40\%$ for M-type stars. The binary fraction is lowest among brown dwarfs ($10-20\%$). Assuming that this trend is not the result of selection effects, this trend is inconsistent with PCP-I and PCP-III (for which $\binf$ is independent of $M_1$) nor SCP-I and SCP-III (for which $\binf$ only mildly increases with $M_1$). Furthermore, observations of the mass ratio distribution have ruled out RP and PCRP (see above) in most stellar populations. Of the pairing functions described in these proceedings, only PCP-II and SCP-II remain viable options. A deeper analysis, including a study of more complicated pairing functions, combined with further deep observations, is necessary for a full description of the pairing function in the different stellar populations.

When deriving a pairing function from observations, a complication is introduced by selection effects. The origin of these selection effects is two-fold: sample biases (introduced by the choice of the observed sample), and instrumental biases (e.g., due to limited spatial resolution or sensitivity); see Kouwenhoven (2006) for details. The instrumental biases often prevent the detection chances of low-mass companion stars in a survey for binarity. Such selection effects are usually well understood and can be taken into account. The selection of the targets in a binarity survey introduces a sample bias. In the sections above we have seen that the mass ratio distribution and binary fraction depend strongly on the choice of the sample. In order to properly constrain the pairing function of a stellar population, a survey of stars with a large range in spectral types is a prerequisite. To obtain an estimate of the {\em primordial} pairing function, one additionally needs to correct for the effects of dynamical and stellar evolution that have altered the binary population over time.

\section{Summary and discussion} \label{section:summary}

We have described several simple algorithms of pairing individual stars into binary systems, and refer to these algorithms as {\em pairing functions} (Kouwenhoven et al. 2008). Each pairing function is characterized by a generating mass distribution $f_M(M)$ and a generating binary fraction $\binf$, and most pairing functions by a generating mass ratio distribution $\fq$. Several pairing functions can result in a binary fraction that differs significantly from the generating binary fraction $\binf$. The major differences are the dependence of the specific mass ratio distribution and the specific binary fraction on the selected primary mass range (see Figs~\ref{fig:massratiodistributions} and~\ref{figure:binaryfraction_versus_spectraltype}). 

The pairing functions in these proceedings are described because of their simplicity or frequent use in literature. The pairing function in real stellar populations could be different. The pairing function resulting from the star formation process may have an environmental dependence, and is altered over time by dynamical interactions between stars. Kroupa (1995a, 1995b, 1995c) and Thies \& Kroupa (2007), for example, find that young binary populations are well described as a mixture of two different binary populations, both of which are randomly paired from restricted mass ranges. Note that such a heterogeneous population resulting from {\em restricted random pairing} (RRP) over limited mass ranges, is significantly different from the population that results from random pairing over the full mass range (RP).

Each pairing function described in this paper results in a different mass ratio distribution and binary fraction. These properties also depend strongly on the targeted mass range in a survey for binarity. In these proceedings we have shown that it is important to carefully study selection effects in observations, and to clearly state the pairing mechanism used in computer simulations, in order to draw conclusions on the overall binary population. The pairing functions described above are used in literature, but may not accurately describe realistic stellar populations. The next step towards understanding the star formation process is to {\em fully} characterize the binary population in young stellar groupings. Variations in the {\em primordial} pairing function (e.g., binary fraction or mass ratio distribution) with environment  necessarily imply variations in the IMF and star formation mechanism.


\acknowledgements

M.K. was supported by PPARC/STFC under grant number PP/D002036/1 and by NWO under grant number 614.041.006. This research was supported by the Netherlands Research School for Astronomy. M.K. wishes to thank the organising committee for a travel grant.


\appendix

\end{document}